\documentclass[twocolumn,amsmath,amssymb,prb,superscriptaddress,nofootinbib]{revtex4}

\pdfoutput=1
\usepackage{upgreek}
\usepackage{bm}
\usepackage{pifont}
\usepackage{graphicx}
\usepackage{color}

\usepackage{ulem}

\begin{document}

\preprint{APS/123-QED}
\title{Disorder-Induced excitation continuum in a spin 1/2 triangular lattice cobaltate}

\author{Bin Gao}
\email{bin.gao@rice.edu}
\thanks{These authors contributed equally to this work.}
\affiliation{Department of Physics and Astronomy, Rice University, Houston, Texas 77005, USA}
\author{Tong Chen}
\thanks{These authors contributed equally to this work.}
\affiliation{Department of Physics and Astronomy, Rice University, Houston, Texas 77005, USA}
\author{Chien-Lung Huang}
\affiliation{Department of Physics and Center for Quantum Frontiers of Research \& Technology (QFort), National Cheng Kung University, Tainan 701, Taiwan}
\author{Yiming Qiu}
\affiliation{NIST Center for Neutron Research, National Institute of Standards and Technology, Gaithersburg, Maryland 20899, USA}
\author{Guangyong Xu}
\affiliation{NIST Center for Neutron Research, National Institute of Standards and Technology, Gaithersburg, Maryland 20899, USA}
\author{Jesse Liebman}
\affiliation{Department of Physics and Astronomy, Rice University, Houston, Texas 77005, USA}
\author{Lebing Chen}
\affiliation{Department of Physics and Astronomy, Rice University, Houston, Texas 77005, USA}
\author{Matthew B. Stone}
\affiliation{Neutron Scattering Division, Oak Ridge National Laboratory, Oak Ridge, Tennessee, 37830, USA}
\author{Erxi Feng}
\affiliation{Neutron Scattering Division, Oak Ridge National Laboratory, Oak Ridge, Tennessee, 37830, USA}
\author{Huibo Cao}
\affiliation{Neutron Scattering Division, Oak Ridge National Laboratory, Oak Ridge, Tennessee, 37830, USA}
\author{Xiaoping Wang}
\affiliation{Neutron Scattering Division, Oak Ridge National Laboratory, Oak Ridge, Tennessee, 37830, USA}
\author{Xianghan Xu}
\affiliation{Rutgers Center for Emergent Materials and Department of Physics and Astronomy,
Rutgers University, Piscataway, New Jersey 08854, USA}
\author{Sang-Wook Cheong}
\affiliation{Rutgers Center for Emergent Materials and Department of Physics and Astronomy,
Rutgers University, Piscataway, New Jersey 08854, USA}
\author{Stephen M. Winter}
\email{winters@wfu.edu}
\affiliation{Department of Physics and Center for Functional Materials, Wake Forest University, NC 27109, USA}
\author{Pengcheng Dai}
\email{pdai@rice.edu}
\affiliation{Department of Physics and Astronomy, Rice University, Houston, Texas 77005, USA}

\begin{abstract} 
 A spin-1/2 triangular-lattice antiferromagnet is a prototypical frustrated
quantum magnet, which exhibits remarkable quantum many-body effects that arise from the synergy between geometric spin frustration and quantum fluctuations. It can host quantum frustrated magnetic topological phenomena like quantum spin liquid (QSL) states, highlighted by the presence of 
fractionalized quasiparticles within a continuum of magnetic excitations. In this work, we use neutron scattering to study CoZnMo$_3$O$_8$, which has a triangular lattice of $J_{\rm eff} = 1/2$ Co$^{2+}$ ions with octahedral coordination. We found a wave-vector dependent excitation continuum at low energy that disappears with increasing temperature.  Although these excitations are reminiscent of a spin excitation continuum in a QSL state, their presence in CoZnMo$_3$O$_8$ originates from magnetic intersite disorder induced dynamic spin states with peculiar excitations. Our results therefore give direct experimental evidence for the presence of a disorder-induced spin excitation continuum.
\end{abstract}
\maketitle

\section{Introduction}
A quantum spin liquid (QSL) is a state of matter in which the spins of unpaired electrons in a solid are quantum entangled, but do not show magnetic order in the zero-temperature limit \cite{Balents2010,Zhou2017,Savary2017,Broholm2020}. Because such a state may be important to the microscopic origin of high-$T_c$ superconductivity \cite{Lee2006} and useful for quantum computation in certain cases \cite{Kitaev2006}, experimental realization of QSLs is a long-sought goal in modern condensed matter physics.  In this context, Spin-1/2 antiferromagnets, especially 2D triangular lattice systems, have attracted great interest due to their intriguing magnetic properties. Several Yb-based and Co-based triangular antiferromagnets have been proposed as candidates for QSL state, such as YbMgGaO$_4$ \cite{Shen2016,Paddison2017}, NaYbSe$_2$ \cite{Dai2021}, and Na$_2$BaCo(PO$_4$)$_2$ \cite{Zhong2019,Li2020}. 
Although INS experiments on the 2D triangular lattice materials YbMgGaO$_4$ and NaYbSe$_2$ have found evidence for the hallmark of a QSL, a continuum of magnetic excitations near the Brillouin zone boundary in the low  temperature limit, one major issue in these candidates is the presence of disorder on either non-magnetic or magnetic sites, like the mixing of Mg and Ga sites in YbMgGaO$_4$ \cite{Ma2018}, $\sim$4\% vacancy of Na site occupied by Yb in NaYbSe$_2$ \cite{Dai2021}, respectively. While Na$_2$BaCo(PO$_4$)$_2$ actually orders antiferromagnetically below 1 K and thus cannot be an ideal QSL \cite{Li2020}, 
the disorder in other 2D triangular lattice QSL candidates complicates the interpretation of the data \cite{Kimchi2018,PhysRevLett.119.157201}. Therefore, it is important to understand how disorder/randomness in a QSL candidate affects its true magnetic ground state. Particularly, in YbMgGaO$_4$, the frequency dependence of AC magnetic susceptibility was observed, indicating a spin-glass ground state instead of a QSL \cite{Ma2018}. In addition, the broad excitation continua observed by INS are believed to arise from the spin-glass ground state, instead of a QSL \cite{Ma2018}.   Despite the intensive study, it remains unresolved as to whether YbMgGaO$_4$ is a QSL or has a random-singlet ground state. A recent study \cite{Rao2021} suggests that a QSL state with itinerant excitations and quantum spin fluctuations survives the disorder in YbMgGaO$_4$.
The ultimate question is whether the QSL survives the disorder or the disorder leads to a ``spin-liquid-like" state, such as the proposed random-singlet state\cite{Kimchi2018,PhysRevLett.119.157201} . Similar discussions have also emerged around the organic triangular lattice QSL candidates \cite{watanabe2014quantum,riedl2019critical}, $1T-$TaS$_2$,\cite{murayama2020effect} and Kitaev materials \cite{li2018role,knolle2019bond,kao2021vacancy,bahrami2021effect}. Since the disorder is unavoidable in most QSL materials, this question represents a major challenge for QSL candidates.

Recently, there has been a great deal of interest in the family of M$_2$Mo$_3$O$_8$ (M = Fe, Mn, Co, Ni and Zn), mainly due to their multiferroic properties \cite{Wang2015a,Kurumaji2017,Yu2018a,Tang2019,Morey2019}. Multiferroic materials, which can possess more than one primary ferroic order in a single phase, have driven significant research trends \cite{Cheong2007a,Dong2015}. The whole series of compounds all have the same space group $P6_3mc$, which belongs to a polar point group \cite{McAlister1983a}. They also contain magnetic ions (Fe, Mn, Co, and Ni), and are therefore polar magnets. Below magnetic ordering temperatures, they should possess breaking of time reversal symmetry and space inversion symmetry simultaneously, thus exhibiting non-trivial ME effects. Indeed, various experiments have been performed on this series of compounds and novel properties have been found, including the large magnetoelectric coupling in Fe$_2$Mo$_3$O$_8$ \cite{Wang2015a,Kurumaji2015} and Mn$_2$Mo$_3$O$_8$ \cite{Kurumaji2017}, the ferromagnetism induced by doping \cite{Nakayama2011} , optical diode effect in (Fe,Zn)$_2$Mo$_3$O$_8$ \cite{Yu2018a}, and other optical properties \cite{Stanislavchuk2020,Sheu2019,Kurumaji2017a,Csizi2020}. The magnetoelectric properties are also studied in Co$_2$Mo$_3$O$_8$ \cite{Tang2019} and Ni$_2$Mo$_3$O$_8$ \cite{Tang2021}. Magnetoelectric coupling has been observed in the single crystalline Co$_2$Mo$_3$O$_8$ but was very weak in the polycrystalline samples \cite{Tang2019}, and it is quite different from other compounds in the M$_2$Mo$_3$O$_8$ family such as Fe$_2$Mo$_3$O$_8$ \cite{Wang2015a}.

Meanwhile, there has been increasing interest in the study of the magnetic properties of $d^7$ Co$^{2+}$ in octahedral crystal field environments due to the possibility of realising strongly anisotropic and bond-dependent interactions \cite{Motome2020,Liu2020a,Liu2018a,sano2018kitaev}. Such ions have nominal electronic configuration $(t_{2g})^5 (e_g)^{2}$, corresponding to $S = 3/2$ and an angular momentum $L = 1$. Due to spin-orbit coupling (SOC), the ground state becomes a $J_{\rm eff} = 1/2$ pseudospin doublet \cite{Liu2018a,sano2018kitaev}. The specific spin-orbital composition of the moments allows for anisotropic magnetic interactions. Although for $3d$ transition metal ions such as Co$^{2+}$, SOC is much weaker compared with e.g. $5d^5$ iridates or $4d^5$ ruthenates, honeycomb $3d^7$ cobaltates have received a lot of attention lately. For example, BaCo$_2$(AsO$_4$)$_2$ \cite{Zhong2020,Zhang2023,Shi2021a,Tu2022,Halloran2023}, Na$_3$Co$_2$SbO$_6$ and Na$_2$Co$_2$TeO$_6$ \cite{Kim2022} with an edge-sharing honeycomb lattice of Co$^{2+}$ ions have been proposed as Kitaev QSLs \cite{Kitaev2006,Takagi2019}. However, recent theoretical \cite{das2021xy,maksimov2022ab,winter2022magnetic} and experimental \cite{Shi2021a,Zhang2023,Halloran2023} works have shown conflicting conclusions on the details of the magnetic interactions, with some proposing that magnetic couplings with uniform $XXZ$ anisotropy provide a more accurate description of these compounds compared to bond-dependent Kitaev interactions. This calls for the examination of more Co$^{2+}$ systems, with a particular focus on a range of bonding geometries apart from the studied edge-sharing compounds. 

In this work, we investigate a new compound with nominal composition CoZnMo$_3$O$_8$. In the M$_2$Mo$_3$O$_8$ family of compounds, there are two different coordination environments for the transition metal M in the pseudo-honeycomb M-O layer: the octahedral ($O_h$) and tetrahedral ($T_d$) sites [Figs. 1(a,b)]. For Co$_2$Mo$_3$O$_8$,
the $T_d$ sites have configuration $(e_g)^4 (t_{2g})^3$, and thus no unquenched angular momentum to zeroth order, leading to pure $S = 3/2$ spins.  
Previous studies in this series of compounds show that the doped nonmagnetic ions (Zn$^{2+}$ or Mg$^{2+}$) into M$_2$Mo$_3$O$_8$ prefer to occupy the tetrahedral sites, for example, in FeZnMo$_3$O$_8$\cite{Varret1972a,Bertrand1975}, NiMgMo$_3$O$_8$ \cite{Morey2019} and Co$_{2-x}$Zn$_x$Mo$_3$O$_8$ ($x$ $\lesssim$ 0.55) \cite{Prodan2022}. This is because octahedral site preference energy, defined as the difference between crystal field stabilization energies of a non-octahedral complex and the octahedral complex, is zero for Zn$^{2+}$ ($d^{10}$) or Mg$^{2+}$ ($d^{0}$). In the ideal case, CoZnMo$_3$O$_8$ should form a perfect triangular lattice with Co$^{2+}$ ions on the octahedral sites with a $J_{\rm eff} = 1/2$ Kramers doublet ground state. However, our single crystal neutron diffraction study reveals a significant ($\sim$18\%) tetrahedral and octahedral intersite mixing of Co ions, resulting in an intriguing broad magnetic response below 1 K reminiscent of the spin excitation continuum expected for a triangular lattice QSL. We extracted the magnetic exchange parameters from fitting the spin excitation spectra and comparing them with the calculation using an $ab$-initio approach from a Heisenberg-Kitaev-$\Gamma$ ($HK\Gamma$) model plus additional $XXZ$ anisotropy. Our results indicate that the broad dispersive spin excitations seen by INS are a manifestation of the spin disorder, arising from local spin excitations of different magnetic exchange couplings from the  tetrahedral and octahedral sites rather than a continuum of fractionalized excitations of a QSL.

\section{Experiments}

\begin{figure}[t]
\includegraphics[width=\columnwidth]{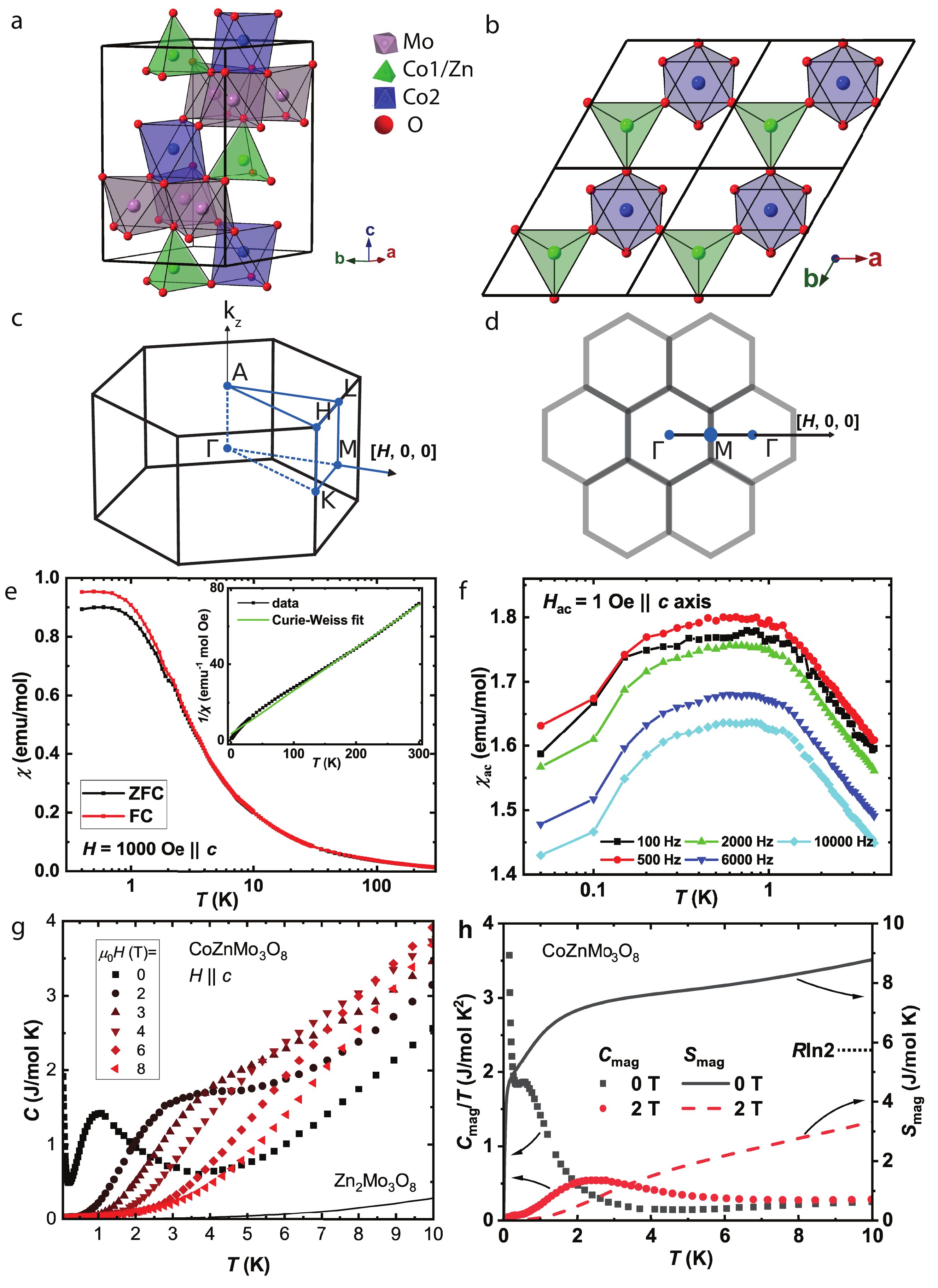}
\caption{
(a,b) Crystal structure of Co$_2$Mo$_3$O$_8$/CoZnMo$_3$O$_8$. The green (blue) ions are the tetrahedral (octahedral)  Co$^{2+}$  and are labeled as Co1/Zn (Co2). The magenta Mo$^{4+}$ ions form non-magnetic trimers. 
(c,d) 3D and 2D reciprocal space of Co$_2$Mo$_3$O$_8$/CoZnMo$_3$O$_8$, where the high symmetry positions are marked. (e) Temperature dependence of magnetic susceptibility under 0.1 T magnetic field along the c-axis. Inset: temperature dependence of the inverse susceptibility and the Curie-Weiss fit in the high-temperature range. 
(f) AC magnetic susceptibility from 0.05 K to 4 K, with a 1 Oe Ac field along the c-axis with a frequency from 100 Hz to 6000 Hz. (g) Temperature dependence of specific heat of CoZnMo$_3$O$_8$ under various magnetic fields along the c-axis. The data for the non-magnetic Zn$_2$Mo$_3$O$_8$ in zero field is also plotted, and used as phonon contribution. 
(h) Left: magnetic contribution to the specific heat $C_{mag}/T$. Right: magnetic entropy $S_{mag}$ under zero field and 2 T field.}
\label{fig_1}
\end{figure}

Polycrystalline Co$_2$Mo$_3$O$_8$, Zn$_2$Mo$_3$O$_8$ and CoZnMo$_3$O$_8$ were prepared using a solid state method. Stoichiometric Co$_3$O$_4$ (99.9\% Alfa Aesar), ZnO (99.9\% Alfa Aesar), Mo (99.9\% Alfa Aesar), and MoO$_3$ (99.9\% Alfa Aesar) powders were mixed, pelleted and sintered at 1050$^{\circ}$C in an evacuated quartz tube. Single crystals were synthesized using the chemical vapor transport method \cite{Strobel1983b}. Typical crystals obtained were 1 mm $\times$ 1 mm $\times$ 0.5 mm, with well-defined hexagonal edges. Laue patterns of single crystal and X-ray powder diffraction patterns of ground powders from crystals show a pure phase with space group $P6_3mc$ and lattice constants $a$ = $b$ = 5.767 \AA \ and $c$ = 9.914 \AA \ for Co$_2$Mo$_3$O$_8$, and $a = b = 5.783$ \AA \ and $c = 9.902$ \AA \ for CoZnMo$_3$O$_8$. The crystal structures consist of pseudo-honeycomb layers of Co$^{2+}$/Zn$^{2+}$, separated by Mo$^{4+}$ layers [Fig.~\ref{fig_1}(a)]. The Mo$^{4+}$ kagome-like layer is trimerized. The Mo trimers are in the singlet state and do not contribute to magnetism \cite{Cotton1964}.

For Co$_2$Mo$_3$O$_8$, prior neutron diffraction experiments \cite{Tang2019} on polycrystalline samples revealed a layered collinear antiferromagnetic magnetic structure below $T_N \approx 40$ K, with moments oriented along the $c$-axis. The large ordering temperature indicates strong antiferromagnetic couplings between nearest neighbor $O_h$ and $T_d$ sites. The refined moments for the two Co sites were quite close, 3.44(1) $\mu_B$ for the tetrahedral  Co$^{2+}$, and 3.35(1) $\mu_B$ for the octahedral Co$^{2+}$ site. These are compatible with the ideal values of 3.87 $\mu_B$ for the $S = 3/2$ moments at the $T_d$ sites (with a $g$-value of $2$), and 3.75 $\mu_B$ for $J_{\rm eff} = 1/2$ moments at the $O_h$ sites (with ideal isotropic $g = 4.33$; see Ref.~\onlinecite{lines1963magnetic}). 

For CoZnMo$_3$O$_8$, the Zn$^{2+}$ ions prefer to occupy the tetrahedral sites. In the ideal case, the remaining Co ions would form a triangular sublattice of $O_h$ sites, as shown in Fig.~\ref{fig_1}(b). In order to probe the composition of our samples, we performed single crystal diffraction at BL-12 TOPAZ\cite{Coates2018}, the neutron time of flight Laue diffractometer at the spallation neutron source (SNS) at the Oak Ridge National Laboratory (ORNL). Details of the refined atomic positions are shown in Table \ref{refinement}. Importantly, we find that the samples are better described by a composition Co$_{1.18}$Zn$_{0.82}$Mo$_3$O$_8$, with roughly 18\% of $T_d$ sites occupied by Co, and from 0\% to 8\% of $O_h$ sites occupied by Zn. As a result of the samples being Co-rich, a fraction of magnetic $T_d$ sites must be considered in the magnetic response, as discussed in detail below. Notice that from Ref.~\onlinecite{Steiner2005} that although compositions of CVT  crystals between different batches may vary, the compositions within each batch stay the same. We used the crystals from the same batch to perform all the experiments including INS.
 
 In principle, this disorder may lead to a spin-glass ground state. To test for potential spin-glass behavior, we measured the AC magnetic susceptibility down to 50 mK [Fig.~\ref{fig_1}(f)]. It shows a broad peak from 0.1 K to 1 K. There is no shift of the peak positions with increasing frequency, indicating no sign of glassy-behavior within the time scale studied in the current work, and no sharp features indicating magnetic order. At higher temperatures, the susceptibility displays Curie-Weiss behavior, as shown in Fig.~\ref{fig_1}(e). The slope of the inverse susceptibility (inset of Fig.~\ref{fig_1}(e)) changes below 50 K, and $\theta _W = -1.5$ K for moderate (1 K $<$ T $<$ 10 K) temperatures, indicating weak antiferromagnetic correlations. 

We also measured the temperature dependence of the heat capacity under several magnetic fields along the $c$-axis [Fig.~\ref{fig_1}(g)]. At zero field, the heat capacity shows a broad bump around 1 K, and the bump shifts to higher temperature with increasing magnetic fields similar to other QSL candidates \cite{Gao2019}. After subtracting the nonmagnetic contribution to the specific heat, Zn$_2$Mo$_3$O$_8$, we extrapolated the data from the lowest temperature $\sim$ 50 mK to absolute zero via a method described in Appendix:D, and integrated $C_{mag}/T$ vs $T$ to calculate the magnetic entropy $S_{mag}$, as shown in Fig. 1(h). At zero field $C_{mag}/T$ shows a fast increase as the temperature decreases below 1~K. This is most probably due to an impurity-induced Schottky anomaly, which additionally enlarges the entropy that $S_{mag}$ reaches $Rln2$ below 1 K.  At higher temperature, the broad hump of $C_{mag}/T$ related to geometric frustration of Co moments adds to the entropy, resulting in $S_{mag} \sim $ 8 J/mol K at 4~K. At $\mu_0H$ = 2 T, the Schottky anomaly is suppressed and a transfer of specific heat weight to high temperature is observed. Regarding the crystal field levels, at the $O_h$ sites the lowest excitations are the local $j_{1/2} \rightarrow j_{3/2}$ excitations, which are expected to appear in the range of 30 meV. Indeed, the first few crystal electric field levels were observed via INS on single crystalline CoZnMo$_3$O$_8$ using $E_i = 40$ meV at the SEQUOIA spectrometer at SNS, ORNL. The powder average $E$ vs $Q$ plot shows several energy levels around 30 meV (see Appendix), consistent with other octahedral Co$^{2+}$ compounds with $J_{\rm eff} = 1/2$ states, like CoTiO$_3$ \cite{Yuan2020a,Elliot2020}, Na$_3$Co$_2$SbO$_6$ and Na$_2$Co$_2$TeO$_6$ \cite{Kim2022}.

\begin{figure}[t]
\includegraphics[width=\columnwidth]{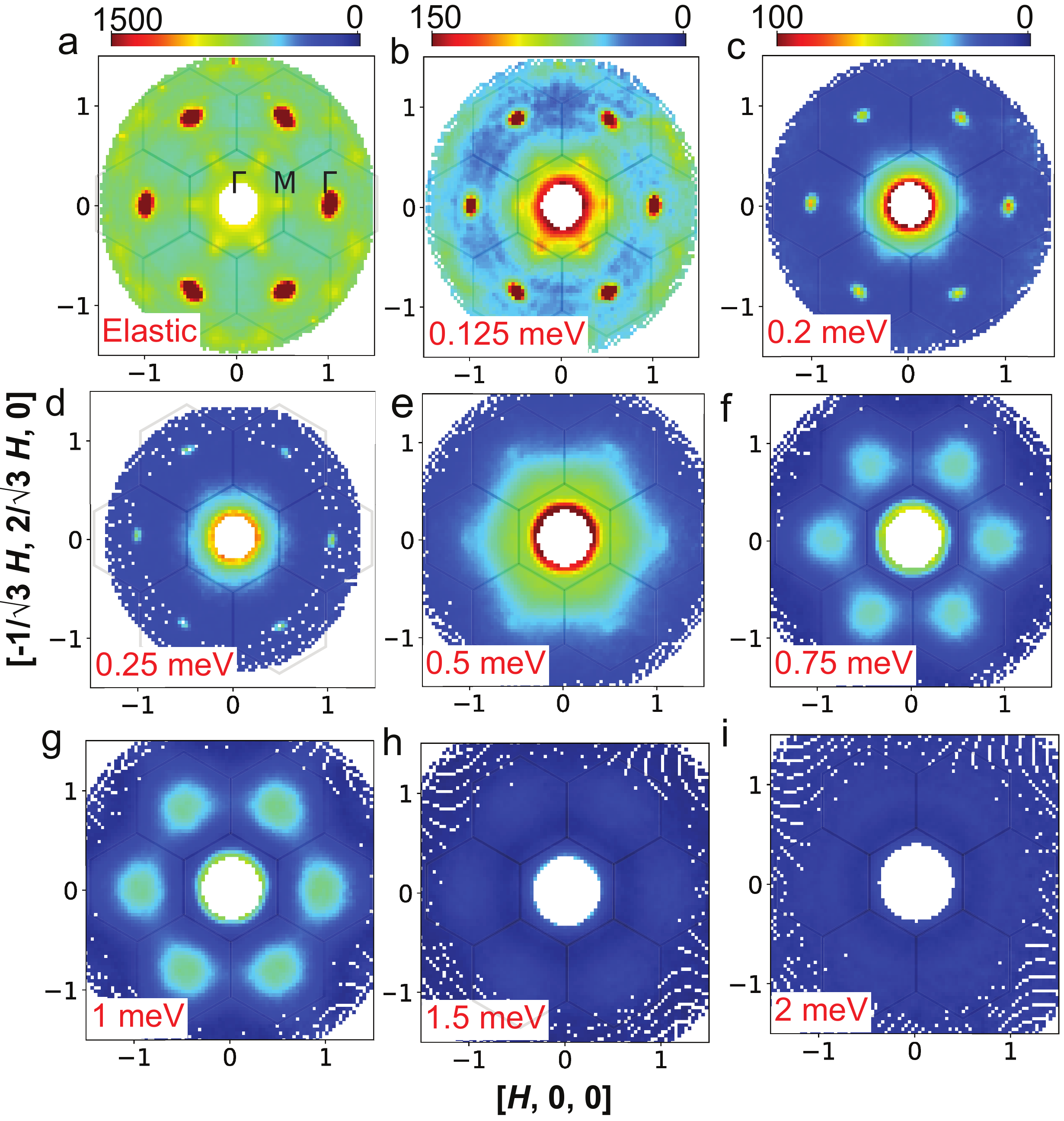}
\caption{Neutron scattering intensity in the $[h,k,0]$ plane at the elastic line (a), E = 0.125 meV (b), E = 0.2 meV (c), E = 0.25 meV (d), E = 0.5 meV (e), E = 0.75 meV (f), E = 1 meV (g), E = 1.5 meV (h), and E = 2 meV (i). The data were collected at T = 0.1 K, scanned with 120$^{\circ}$ rotation, and folded to the whole scattering plane. The sharp peaks at $\Gamma$ points up to E = 0.25 meV are from structure Bragg peaks. Colour bars indicate scattering intensity in arbitrary units (a.u.), and panels c-i use the same color bar.
\label{fig_neutron1}
}
\end{figure}


In order to probe the magnetic ground state and excitations of CoZnMo$_3$O$_8$, we performed neutron scattering experiments on 50 small co-aligned single crystals (about 2 grams) in the $[H, K, 0]$ plane. Scattering experiments were carried out at the multi-axis crystal spectrometer (MACS) \cite{Rodriguez2008} at the NIST Center for Neutron Research (NCNR), National Institute of Science and Technology (NIST). Instrumental energy resolutions at the elastic line are $\Delta E\approx 0.14$ meV for final neutron energy of $E_f = 3.7$ meV. The measured wavevector ($\bf Q$) dependence of  magnetic scattering function $S({\bf Q},E)$ at various energy transfers ($E$) from 0 to 2 meV are shown in Fig.~\ref{fig_neutron1}. Magnetic peaks are observed at the $M$ points at the elastic line, which we discuss below as a characteristic of short-range stripy order. They are still visible at $E = 0.125$ meV due to finite energy resolution. The most striking feature of the response is a continuum of excitations from $E = 0.5$ meV centered around the $M$-points, and evolving to larger wave vectors with increasing energy tranfer. The continuum is most clear at $E = 0.75$ meV and 1 meV, centered around the $\Gamma$ points. Above $E = 1.5$ meV, the continuum scattering weakens and broadens.

\begin{figure}[t]
\includegraphics[width=\columnwidth]{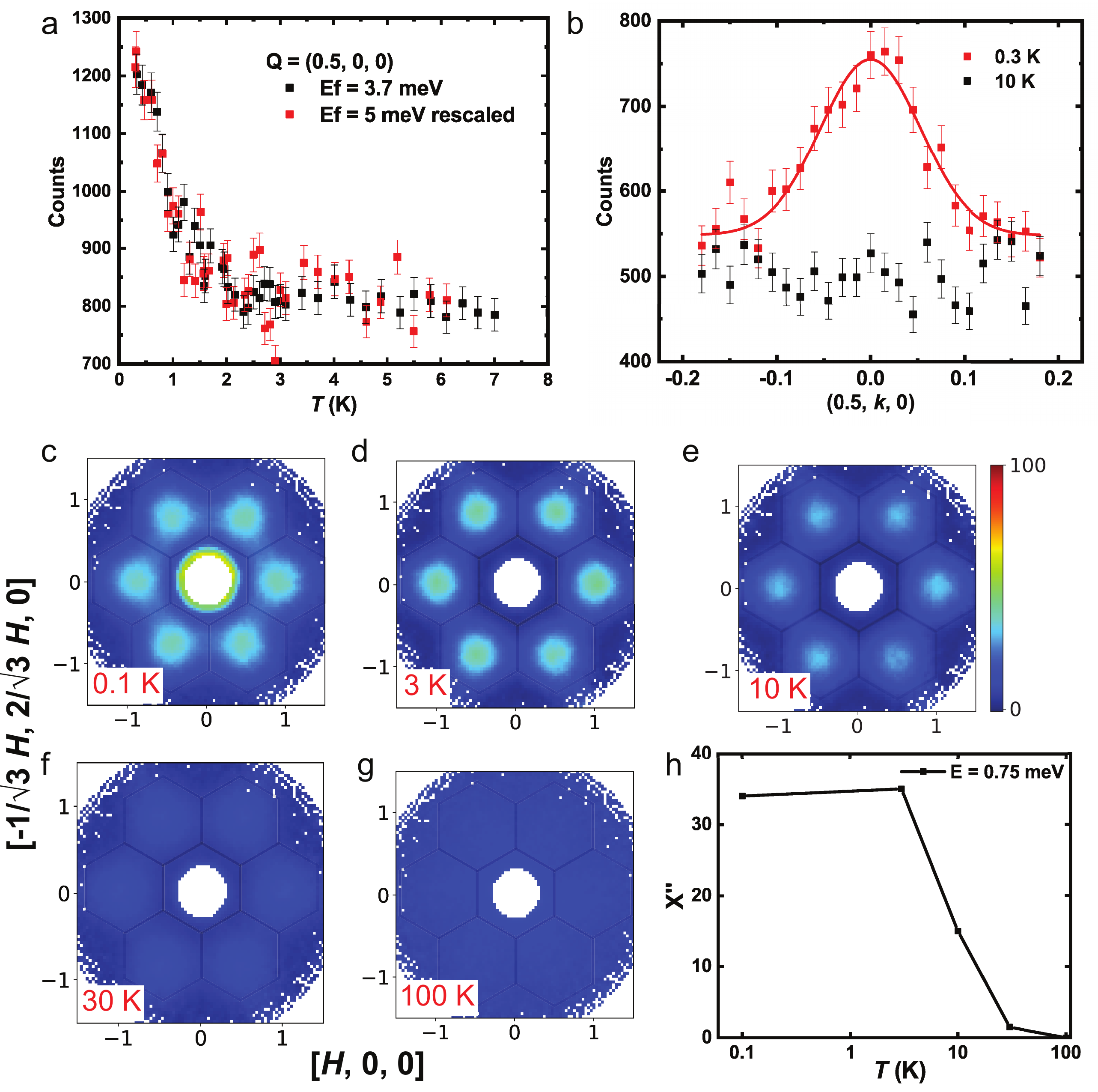}
\caption{The temperature dependence of the magnetic peaks at the elastic line and the continuum from neutron scattering at E = 0.75 meV. (a) Neutron counts at M point Q = (0.5, 0, 0) at the elastic line using two different energies $E_f$ = 3.7 and $E_f$ = 5. The counts using $E_f$ = 5 meV are re-scaled to match the background at high temperature. (b) Q-scan of the magnetic peak at Q = (0.5, 0, 0) at 0.3 K and 10 K. Red solid curve line is the Gaussian fit for the scan at 0.3 K. Error bars represent one standard deviation of statistical uncertainty. (c)-(g) The wavevector dependence of the neutron scattering in the $[h, k, 0]$ plane at $E$ = 0.75 meV at (c) 0.1 K (d) 3 K, (e) 10 K, (f)30 K, (g) 100 K. (h) temperature dependence of the imaginary part of the susceptibility at E = 0.75 meV, by integrating the intensity of the continuum at different temperatures. Colour bars indicate scattering intensity in arbitrary units (a.u.). 
\label{fig_neutron2}
}
\end{figure}

In order to further elaborate on the ground state, we measured the temperature dependence of the intensity at $M$ point, as well as the wavevector dependence of the intensity around $M$ points at the Spin Polarized Inelastic Neutron Spectrometer (SPINS), NCNR using neutron final energies of $E_f = 3.7$ meV and $E_f = 5$ meV (Fig.~\ref{fig_neutron2}). The intensity of the peak at $M$ points drops quickly above 1 K, consistent with the susceptibility measurements. Also, the two data sets using $E_f$ = 3.7 meV and $E_f$ = 5 meV with different energy resolutions are almost identical [Fig. 3(a)]. If the system exhibits a spin-glass state, then the apparent spin-glass transition temperatures would be different dependent on the instrumental  energy-resolution, arising from the different timescale spin dynamics probed with different energy resolution in neutron diffraction measurements as seen in spin-glass order of Fe$_{1+y}$Te$_{1-x}$Se$_x$ \cite{Tian_2021}. From the ${\bf Q}$-scan of the magnetic peak around the $(0.5, K, 0)$ at 0.3 K with $E_f =3.7$ meV (Fig.~\ref{fig_neutron2}(b)), we obtain a full-width-at-half-maximum FWHM = 0.1254 \AA$^{-1}$ from the Gaussian fit. This corresponds to a spin-spin correlation length of $\sim$18 \AA, indicating short range ordering on the scale of several unit cells.

In Fig.~\ref{fig_neutron2}(c-g), we show the temperature dependence of the $\Gamma$-point continuum. The transfer energy was fixed at 0.75 meV, and the temperatures were changed from 0.1 K to 3 K, 10 K, 30 K and 100 K. The shape and position of the continuum remain constant. However, the intensity (after considering the Bose factor) begins to drop only above 3 K, as summarized in Fig.~\ref{fig_neutron2}(h). Thus, the continuum persists to much higher temperatures than the onset of stripe antiferromagnetic correlations.

\section{Theory}
\subsection{Magnetic Interactions}

In both Co$_2$Mo$_3$O$_8$ and CoZnMo$_3$O$_8$ compounds, the magnetic Co$^{2+}$ ions may occupy either tetrahedral ($T_d$) or octahedral ($O_h$) sites. In the former case, the local site Hamiltonian includes single-ion anisotropy $A_c$:
\begin{align}
\mathcal{H}_{T_d} = \sum_n A_c (S_n^c)^2
\label{eqn_Htd}
\end{align}
where $S_n^c$ refers to the component of the spin of the $n$th $T_d$ site along the crystallographic $c$-axis. The interactions between $O_h$ and $T_d$ sites may be generally written as:
\begin{align}
\mathcal{H}_{T_d-O_h} = \sum_{i,n} \mathbf{s}_i \cdot \mathbb{J}_{i,n} \cdot \mathbf{S}_n
\end{align}
where $\mathbf{s}_i$ refers to the spin at the $i$th $O_h$ site. Similarly, interactions between $O_h$ may be written:
\begin{align}
\mathcal{H}_{O_h-O_h} = \sum_{i,j} \mathbf{s}_i \cdot \mathbb{J}_{i,j} \cdot \mathbf{s}_j
\label{eqn_HOh}
\end{align}
Due to the spin-orbital nature of the $O_h$ moments, and the relatively low symmetry of each bond, the magnetic coupling tensors $\mathbb{J}$ may be strongly anisotropic and bond-dependent. In general, there are two interrelated sources of exchange anisotropy \cite{winter2022magnetic}: (i) the underlying couplings, which may take the Heisenberg-Kitaev-$\Gamma$ form for undistorted octahedra and specific bonding arrangements \cite{Liu2018a,sano2018kitaev,das2021xy}, and (ii) additional uniaxial anisotropy imposed by the effects of trigonal distortion on the local $j_{1/2}$ moments \cite{lines1963magnetic,buyers1971excitations}. To capture both effects with minimal parameters, we propose a model including $JK\Gamma$ couplings between $O_h$ sites in the cubic $xyz$ coordinate system modified by additional $XXZ$ anisotropy in the crystallographic $ab^*c$ coordinate system. The latter is imposed by taking $S_i^{a*} \to \eta S_i^{a*}$ and $S_i^{b*} \to \eta S_i^{b*}$. The resulting model can be mapped to $J-K-\Gamma-\Gamma^\prime$ couplings in the cubic coordinates:
\begin{figure}[b]
\includegraphics[width=\columnwidth]{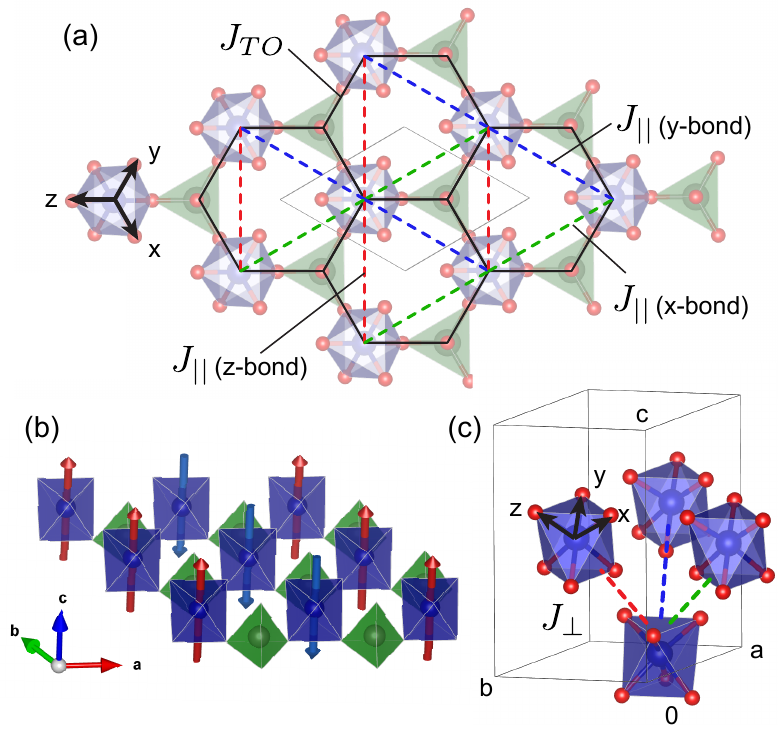}
\caption{Summary of magnetic interactions. (a) intraplane couplings with cubic $(x,y,z)$ coordinates indicated. (b) stripy antiferromagnetic ordering pattern for ideal CoZnMo$_3$O$_8$. (c) interplane couplings $J_\perp$ (approximated as bond-independent). 
\label{fig_interactions}
}
\end{figure}

\begin{align}
\mathcal{H}_{O_h-O_h} =& \sum_{\langle ij \rangle}  \tilde{J} \  \mathbf{s}_i \cdot \mathbf{s}_j + \tilde{K} \ s_i^\gamma s_j^\gamma + \tilde{\Gamma} (s_i^\alpha s_j^\beta + s_i^\beta s_j^\alpha) 
\nonumber \\
& \ + \tilde{\Gamma}^\prime (s_i^\alpha s_j^\gamma + s_i^\gamma s_j^\alpha+s_i^\beta s_j^\gamma + s_i^\gamma s_j^\beta) 
\end{align}
where $(\alpha,\beta,\gamma) = (x,y,z)$ for the Z-bonds, $(y,z,x)$ for the X-bonds, and $(z,x,y)$ for the Y-bonds depicted in Fig.~\ref{fig_interactions}. These couplings can be written as:
\begin{align}
\tilde{J} = & \ \frac{1}{9} \left[(3J+K+2\Gamma) - 2(K-\Gamma)\eta \right. \nonumber \\
 & \hspace{25mm}\left.+(6J+K-4\Gamma)\eta^2\right]
\\
\tilde{K} = & \ \frac{1}{3} \left[2(K-\Gamma)\eta +(K+\Gamma)\eta^2\right]
\\
\tilde{\Gamma} =& \  \frac{1}{9}\left[  (3J + K +2\Gamma) -2(K-\Gamma)\eta \right. \nonumber \\
 & \hspace{25mm}\left.- (3J - 5\Gamma -  K)\eta^2 \right]
\\
\tilde{\Gamma}^\prime =& \  \frac{1}{9} \left[ (3J+K+2\Gamma) +(K-\Gamma)\eta \right. \nonumber \\
 & \hspace{25mm}\left.-(3J+2K+\Gamma)\eta^2\right]
\end{align}
where $J, K,\Gamma$ parameterize the underlying model, and $\eta$ denotes the degree of XXZ anisotropy ($\eta = 0$ refers to pure collinear Ising couplings along the crystallographic $c$-axis, and $\eta =1$ refers to the pure $JK\Gamma$ model). Although the $T_d-O_h$ couplings are also bond-dependent, for simplicity we model them with a pure XXZ Hamiltonian (i.e. $K_{TO} = \Gamma_{TO} = 0$) with the same value of $\eta$, which is equivalent to:
\begin{align}
\mathcal{H}_{T_d-O_h} = \sum_{n,i} J_{TO} \left[  s_i^c S_n^c + \eta \ ( s_i^a S_n^a + s_i^{b*} S_n^{b*})\right]
\end{align}

\begin{table}[t]
\caption {\label{tab_models} Summary of interactions for Co$_2$Mo$_3$O$_8$ and CoZnMo$_3$O$_8$ in meV. We assume both compounds are described by the same set of couplings. The refined model represents a modification to better agree with the reported spin-wave dispersion in Ref.~\onlinecite{reschke2022confirming}. }
\centering\def\arraystretch{1.0}\small
\begin{ruledtabular}
\begin{tabular}{l|ccccccccc}
model& $J_{TO}$ & $A_c$ & $J_{||}$ & $K_{||}$ & $\Gamma_{||}$ & $J_\perp$ & $K_\perp$& $\Gamma_\perp$ & $\eta$
\\
\hline
{\it ab-initio} & 3.3 & -0.2 & 0.53 & -0.03& 0.05& 0.3 & 0.1 &-0.2& 0.5-0.8
\\
refined & {\bf 2.9} & {\bf -0.5} & 0.53 & -0.03& 0.05 & 0.3 & 0.1 &-0.2& {\bf 0.7}
\end{tabular}
\end{ruledtabular}
\end{table}

In order to estimate the couplings computationally, we first computed fully relativistic hoppings between $d$-orbitals with FPLO \cite{eschrig2004relativistic} at the GGA (PBE) level. For this purpose, we employed the reported single-crystal structure of Co$_2$Mo$_3$O$_8$ of Ref.~\onlinecite{Tang2019}. The resulting hoppings were then employed in exact diagonalization calculations for one or two Co sites, with local Coulomb interactions modelled in the spherically symmetric approximation \cite{pavarini2014electronic} with $J_{H} (t_{2g}) = 0.7$ eV, and $U(t_{2g}) = 3.25$ eV, following Ref.~\onlinecite{das2021xy}. Full details including the full $\mathbb{J}$ tensors are given in the Appendix. 

For Co$_2$Mo$_3$O$_8$, the computed couplings are dominated by the nearest neighbor $T_d-O_h$ interactions, which we estimate computationally to be $J_{TO} \sim +3.3$ meV, with $\eta \sim 0.7$, indicating Ising anisotropy. For the single-ion anisotropy of the tetrahedral sites, we compute $A_c \sim -0.2$ meV. Together, the fact that $\eta < 1$ and $A_c < 0$ explains the observed $c$-axis orientation of the ordered moments. In this compound, the ``second neighbor'' $O_h - O_h$ couplings represent subleading interactions, and come in two varieties, as shown in Fig.~\ref{fig_interactions}. The interplane interactions are found to be marginally bond-dependent, with computed values $J_{||} = 0.53$,  $K_{||} = -0.03$ and $\Gamma_{||} = 0.05$ meV (and $\eta = 0.7$).
The interplane interactions are also found to be bond-dependent with $J_\perp = 0.3$, $K_\perp = 0.1$ and $\Gamma_\perp = -0.2$ meV. These values serve as a suitable starting point for analysis of the experimental response. To this end, we have considered the zero-field spin-wave dispersion for Co$_2$Mo$_3$O$_8$ reported in Ref.~\onlinecite{reschke2022confirming}, and find that the experimental data may be suitably reproduced with the ``refined'' parameters in Table \ref{tab_models}, which are guided by the {\it ab-initio} results. In particular, we find that a slightly smaller $J_{TO}$ and enhanced $A_c$ better reproduce the band along the [$H$,0,0] direction. Henceforth, we assume that the same couplings are suitable for CoZnMo$_3$O$_8$, and employ the refined model in all subsequent analyses.

\begin{figure*}[t]
\includegraphics[width=0.9\linewidth]{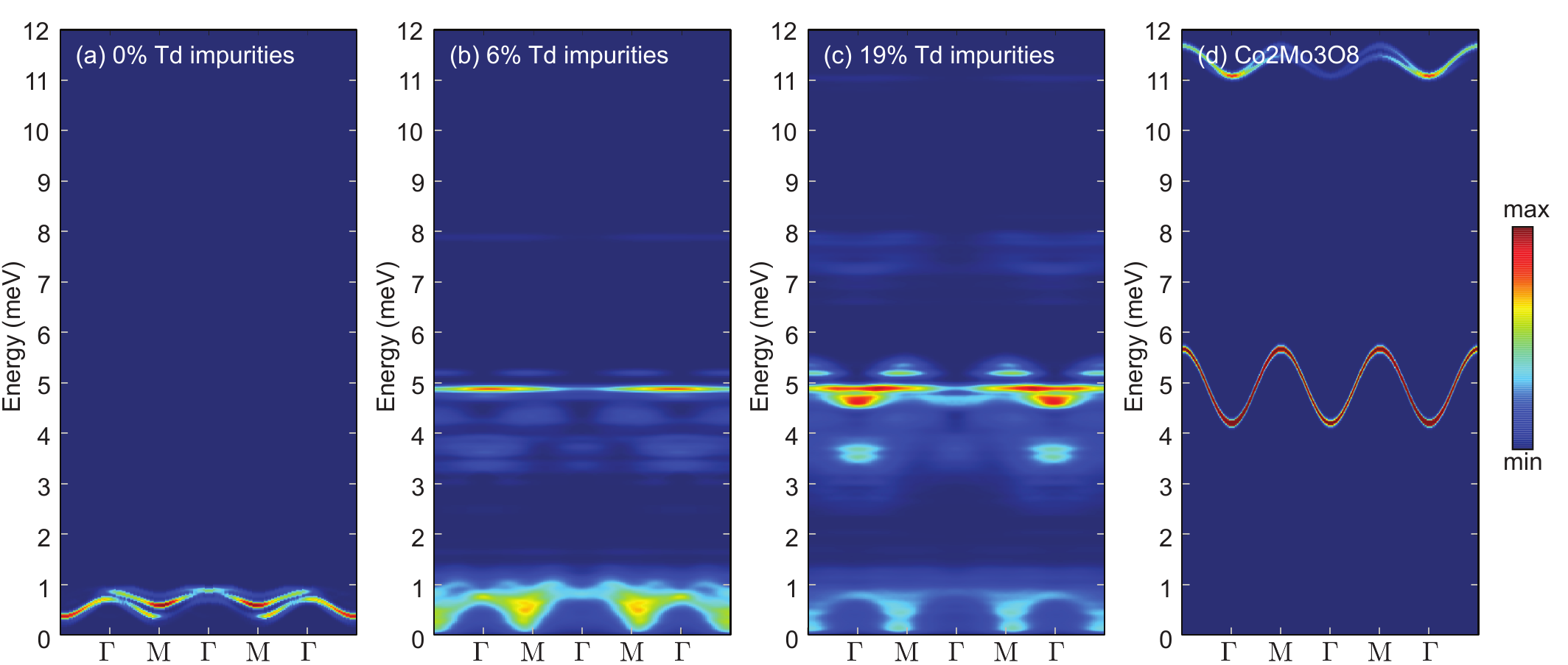}
\caption{Neutron scattering intensity computed by linear spin wave theory for $4 \times 4 \times 1$ supercells with varying $T_d$ site occupancy. (a) Ideal CoZnMo$_3$O$_8$ with no $T_d$ occupancy. (b) With 6\% of $T_d$ sites occupied with Co, corresponding to Co$_{1.06}$Zn$_{0.94}$Mo$_3$O$_8$. (c) With 19\% of $T_d$ sites occupied with Co, corresponding to Co$_{1.19}$Zn$_{0.81}$Mo$_3$O$_8$. (d) Pure Co$_2$Mo$_3$O$_8$ with $T_d$ sites fully occupied by Co. 
\label{fig_LSWT}
}
\end{figure*}

For ideal CoZnMo$_3$O$_8$ without the intersite disorder, the fact that interplane and intraplane couplings are of a similar order of magnitude implies a 3D magnetic lattice with the connectivity of a hexagonal closed packed lattice. The classical ground state for the refined magnetic couplings is a stripy phase within the plane, as depicted in Fig.~\ref{fig_interactions}(b). The in-plane ordering wavevector is $\vec{Q} = M$, which is consistent with the short-range correlations observed experimentally. At the classical level, the presence of magnetic $T_d$ impurities frustrates the underlying stripy order, because the large antiferromagnetic $J_{TO}$ tends to impose {\it ferromagnetic} correlations between nearby $O_h$ moments. 

\subsection{Neutron Scattering Response}

Given that the $\Gamma$-point continuum response in the vicinity of $E = 1$ meV persists to temperatures well above the onset of stripy correlations and the scale of $J_{||}$, it is reasonable to attribute the effect (in part) to short-range correlations between impurity $T_d$ and nearest neighbor $O_h$ sites. In order to check the effect of $T_d$ Co impurities on the bulk response, we first computed the linear spin wave spectrum for various impurity densities. In each case, we employ a $4\times 4 \times 1$ supercell and average the response over 100 different randomly selected placements of the $T_d$ Co atoms. We also averaged over domains related by 3-fold rotation. We did not consider Zn occupation of the $O_h$ sites. We assume an isotropic $g_{T_d} = 2$ for the $T_d$ sites, and an anisotropic $g_{O_h}^{c} = 6, g_{O_h}^{ab} = 3$ for the $O_h$ sites (according to the crystal field estimated via DFT) \cite{lines1963magnetic}. It should be emphasized that this approach captures only the one-magnon ``transverse'' excitations, which corresponds to $\Delta S_c \approx \pm 1$, given that the ordered moments are always oriented nearly along the $c$-axis. Longitudinal $\Delta S_c \approx 0$ excitations are discussed further below.
 Results are shown in Fig.~\ref{fig_LSWT}. 

For ideal CoZnMo$_3$O$_8$ [Fig.~\ref{fig_LSWT}(a)], the spin waves appear below $\sim 1$ meV, with a minimum in the dispersion appearing near the $M$-points. This result can be anticipated from the fact that the stripy phase is ``Klein-dual''  \cite{chaloupka2015hidden,kimchi2014kitaev} to the ferromagnetic phase on the triangular lattice. With the increasing number of impurities [Fig.~\ref{fig_LSWT}(b,c)], these sharp modes become increasingly broadened and washed out. The response below 1 meV retains some features of the spin waves, showing intensity shifting from $M$ to $\Gamma$ with increasing energy. For small concentrations of $T_d$ impurities (6\%), localized modes already appear in the range of 3-5 meV.
Intensity is mainly centered around the $\Gamma$-points. For larger concentrations (19\%) consistent with the experimental composition analysis, the $T_d$ impurities are sufficiently close to each other that the localized excitations start showing significant dispersion, with intensity shifting from the $\Gamma$-points near 3 meV to the zone boundaries near 5 meV. This dispersion mimics the lowest band in Co$_2$Mo$_3$O$_8$, shown in Fig.~\ref{fig_LSWT}(d). Due to the different spin projections for the $O_h$ and $T_d$ sites, this lowest band in Co$_2$Mo$_3$O$_8$ is primarily the excitations of the $T_d$ spins. 

\begin{figure}[t]
\includegraphics[width=\columnwidth]{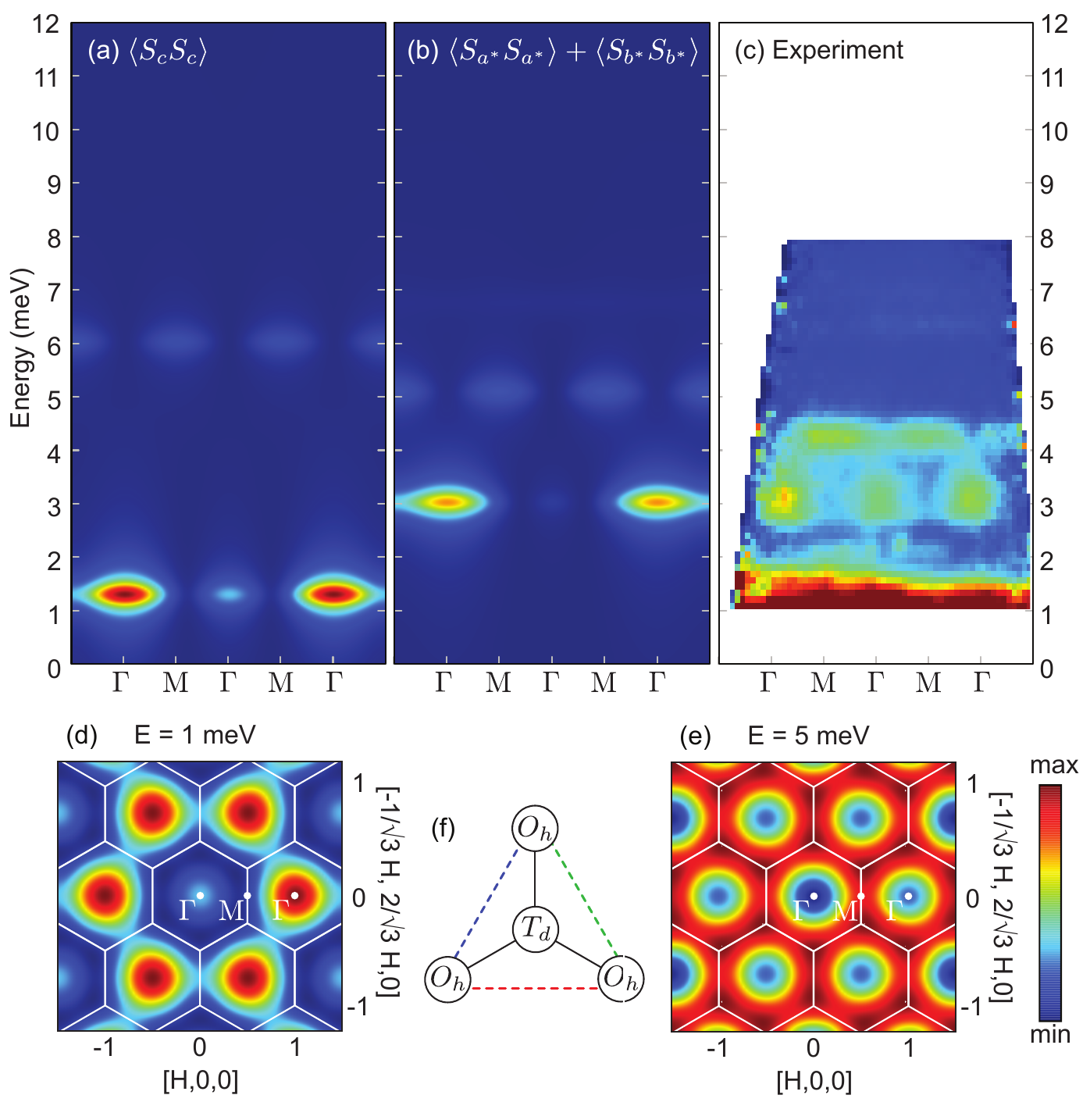}
\caption{Dynamical spin structure factor for the isolated 4-site cluster in the vicinity of a $T_d$ impurity obtained by exact diagonalization. (a) longitudinal $\langle S_c S_c\rangle$ component (b) transverse $\langle S_{a^*} S_{a^*}\rangle+\langle S_{b^*} S_{b^*}\rangle$ component. (c) experimental neutron spectrum plotted along the [$H$, 0, 0] direction. (d,e) total theoretical structure factor evaluated at $E = 1$ and 5 meV. (f) 4-site cluster. Color scales are independent for each figure.
\label{fig_ED}
}
\end{figure}

To capture strong fluctuations around the $T_d$ impurities beyond the linear spin wave approximation (LSWT), we also computed the dynamical spin structure factor via exact diagonalization of a small cluster of four sites shown in Fig.~\ref{fig_ED}(d). For this purpose, we weight the response according to the local $g$-tensors to make the connection with the neutron scattering experiments. The ground state of the cluster is essentially a singlet state with the $S = 3/2$ moments at the central $T_d$ site oppositely aligned with the three $j_{1/2}$ moments of the $O_h$ sites. The lowest excitation appears in the range of 1 meV, and corresponds to an excited $S_c = 0$ triplet, which thus appears in the ``longitudinal'' $\langle S_c S_c\rangle$ channel shown in Fig.~\ref{fig_ED}(a). In the semiclassical (or Ising) limit, this mode would be driven to zero energy, but remains at finite energy provided the spins remain strongly fluctuating in the vicinity of such impurities. The transverse cluster excitations shown in Fig.~\ref{fig_ED}(b) appear in the range of 3-5 meV, and closely resemble the excitations found in LSWT e.g.~Fig.~\ref{fig_LSWT}(c). As shown in Fig.~\ref{fig_ED}(d), the momentum dependence of the scattering intensity  in the range of 1 meV is also consistent with the broad continuum depicted in Fig.~\ref{fig_neutron1} and \ref{fig_neutron2}. For this reason, we attribute such excitations to local impurity-induced scattering intensity. At higher energies, the intensity migrates to the zone edge, as shown in Fig.~\ref{fig_ED}(e).

Turning back to the experiment, in Fig.~\ref{fig_ED}(c), we show the higher energy experimental neutron scattering response for CoZnMo$_3$O$_8$. Data was collected at SEQUOIA, ORNL, using $E_i = 12$ meV at 240 mK. Excitations are clearly observed that are consistent with a significant amount of disorder; a broad continuum with intensity near the $\Gamma$-points at 3 meV disperses to the zone boundary at higher energy. Thus, the consideration of disorder provides a comprehensive explanation of the observed inelastic continuum response. 

\section{Discussion}
In the absence of intersite disorder, CoZnMo$_3$O$_8$ would be a moderately frustrated $XXZ$ antiferromagnet on the triangular lattice, which ultimately orders in a stripe pattern due to additional bond-dependent couplings and interlayer interactions. The presence of a significant density of magnetic Co$^{2+}$ impurities occupying tetrahedral sites competes with this magnetic order. This effect is sufficiently strong to suppress magnetic order, with the experimental correlation length only reaching a few unit cells. Beyond the classical level, the local clusters of $S = 3/2$ $T_d$ sites and three neighboring $J_{\rm eff} = 1/2$ sites form local four-site singlet ground states, implying strong fluctuations of the local moments. This is ensured because the coupling between $T_d$ and $O_h$ sites ($J_{TO}$) greatly exceeds the other couplings, and thus always dominates the local response. The fluctuations are evidently sufficiently strong that a spin-glass state is avoided. Thus, the different energy scales observed in specific heat and neutron scattering can be fully understood: correlations between $T_d$ and $O_h$ spins are fully developed around 10 K, while stripy antiferromagnetic correlations between remaining $O_h$ spins onset only below 1 K. 

The four-site singlets have specific signatures in the dynamical response: a ``longitudinal'' $\Delta S_c = 0$ excitation near 1 meV, with broad intensity peaked at the $\Gamma$-points, and ``transverse'' $\Delta S_c = \pm 1$ excitations in the range 3 - 5 meV, which may appear to disperse from $\Gamma$ to the Brillouin zone edge (particularly for higher impurity concentrations). These are indeed observed experimentally. 

In recent years, the disorder effects in YbMgGaO$_4$ and YbZnGaO$_4$ have been thoroughly studied. A hypothetical, disorder-free version of YbMgGaO$_4$ should exhibit a robust stripe magnetic order as well\cite{Kimchi2018,PhysRevLett.119.157201}, same as the disorder-free  CoZnMo$_3$O$_8$. Examples of quenched disorder include impurities randomly distributed in a crystal lattice or random defects in the arrangement of atoms in a material. The quenched, spatially fluctuating charge environment of the magnetic Yb$^{3+}$ ions due to random occupancy of Mg$^{2+}$ and Ga$^{3+}$ sites is widely believed to be the cause, affecting the low-energy effective spin Hamiltonian through the spin-orbit coupling. Quenched disorder, such as impurities or defects, introduces local variations in the magnetic interactions, leading to frustration and disorder in the overall spin arrangement, and the formation of local ordered moments or spin-glass-like behavior. This can inhibit the formation of long-range spin correlations and destabilize the QSL state.

While in CoZnMo$_3$O$_8$, we are studying a different phenomenology, i.e. positional disorder. The positional disorder can alter the exchange paths and strengths between neighboring spins. The effect of positional disorder due to thermal vibrations can have various consequences. It can broaden the diffraction peaks in X-ray or neutron scattering experiments, leading to a decrease in the sharpness of diffraction patterns (see Fig.\ref{fig_neutron2}(b)). This disruption can hinder the formation of the QSL state, as the precise arrangement and interactions of spins are crucial for the emergence of exotic magnetic behaviors.

In both cases, the disorder can have detrimental effects on the stability and properties of QSL candidates. It can hinder the formation of long-range spin correlations, introduce local magnetic moments, suppress the emergence of gapless excitations, or induce magnetic order. Therefore, when studying QSLs one needs to carefully characterize and control the disorder in materials to better understand and explore their quantum behavior.

Taken together, CoZnMo$_3$O$_8$ serves as an example where experimental signatures typically associated with quantum spin liquids may manifest from disorder. For example, a wide separation of thermodynamic energy scales arises here due to different $J$-values, rather than frustration. The broad, dispersive inelastic response is due to essentially local excitations, rather than a continuum of fractionalized excitations.

\begin{acknowledgments}
The INS work at Rice is supported by the US DOE, BES under grant no. DE-SC0012311 (P.D.).
The single crystal growth and characterization efforts at Rice are supported by the Robert A. Welch Foundation grant nos C-1839 (P.D.). A portion of this research used resources at the Spallation Neutron
Source, a DOE Office of Science User Facility operated by ORNL.
We thank Dr. M.-K. Lee and C.-C. Yang at PPMS-16T and SQUID VSM Labs, Instrumentation Center, National Cheng Kung University (NCKU) for technical support. CLH is supported by the Ministry of Science and Technology in Taiwan (Grants No. MOST 109-2112-M-006-026-MY3 and No. MOST 110-2124-M-006-009).
Access to MACS was provided by the Center for High Resolution Neutron Scattering, a partnership between the National Institute of Standards and Technology and the National Science Foundation under Agreement No. DMR-1508249.
EF and HC acknowledge support from the US Department of Energy (DOE), Office of Science, Office of Basic Energy Sciences, Early Career Research Program Award KC0402020, under Contract No. DE-AC05-00OR22725.
The work at Rutgers University was supported by the DOE under Grant No. DOE: DE-FG02-07ER46382. 
The identification of any commercial product or trade name does not imply endorsement or recommendation by the National Institute of Standards and Technology. 
DFT calculations were performed using the Wake Forest University (WFU) High Performance Computing Facility, a centrally managed computational resource available to WFU researchers including faculty, staff, students, and collaborators \cite{WakeHPC}. 
\end{acknowledgments}

\appendix

\section{Crystal Structure}
As discussed in the main text, the CoZnMo$_3$O$_8$ samples were measured via single crystal diffraction at SNS BL-12 TOPAZ, ORNL, the neutron Time of Flight Laue diffractometer. 4071 peaks (496 non-equivalent peaks) were collected and used to refine the structure. Neutron absorption correction has been performed with transmission factor: $T_{min}$ = 0.888 and $T_{max}$ = 0.975.
The model used is Site1: Zn doped with Co and Site2: Co doped with Zn.

\begin{table}[h]
\caption {neutron diffraction refinement} \label{refinement} 
\begin{tabular}{llllllllll}
Atom & Occ   & X      & Y      & Z      & U \\
Co1    & 0.233 & 0.3333 & 0.6667& 0.95057 & 0.006 \\
Co2     & 0.942  & 0.3333 & 0.6667 & 0.51355 & 0.006  \\
Zn1     & 0.767 & 0.3333 & 0.6667 & 0.95057 & 0.006  \\
Zn2     & 0.058 & 0.3333 & 0.6667 & 0.51355 & 0.006  \\
Mo1     & 1.000      & 0.14571  & 0.85429  & 0.25156  & 0.004  \\
O1      & 1.000      & 0      & 0      & 0.39348 & 0.006  \\
O2      & 1.000      & 0.3333 & 0.6667 & 0.14764 & 0.006\\
O3     & 1.000      & 0.48770 & 0.51230 & 0.36721 & 0.006 \\
O4      & 1.000      & 0.16712 & 0.83288 & 0.63576 & 0.007 
\end{tabular}
\end{table}

\section{Crystal Field Excitations}
\begin{figure}[t]
\includegraphics[width=\columnwidth]{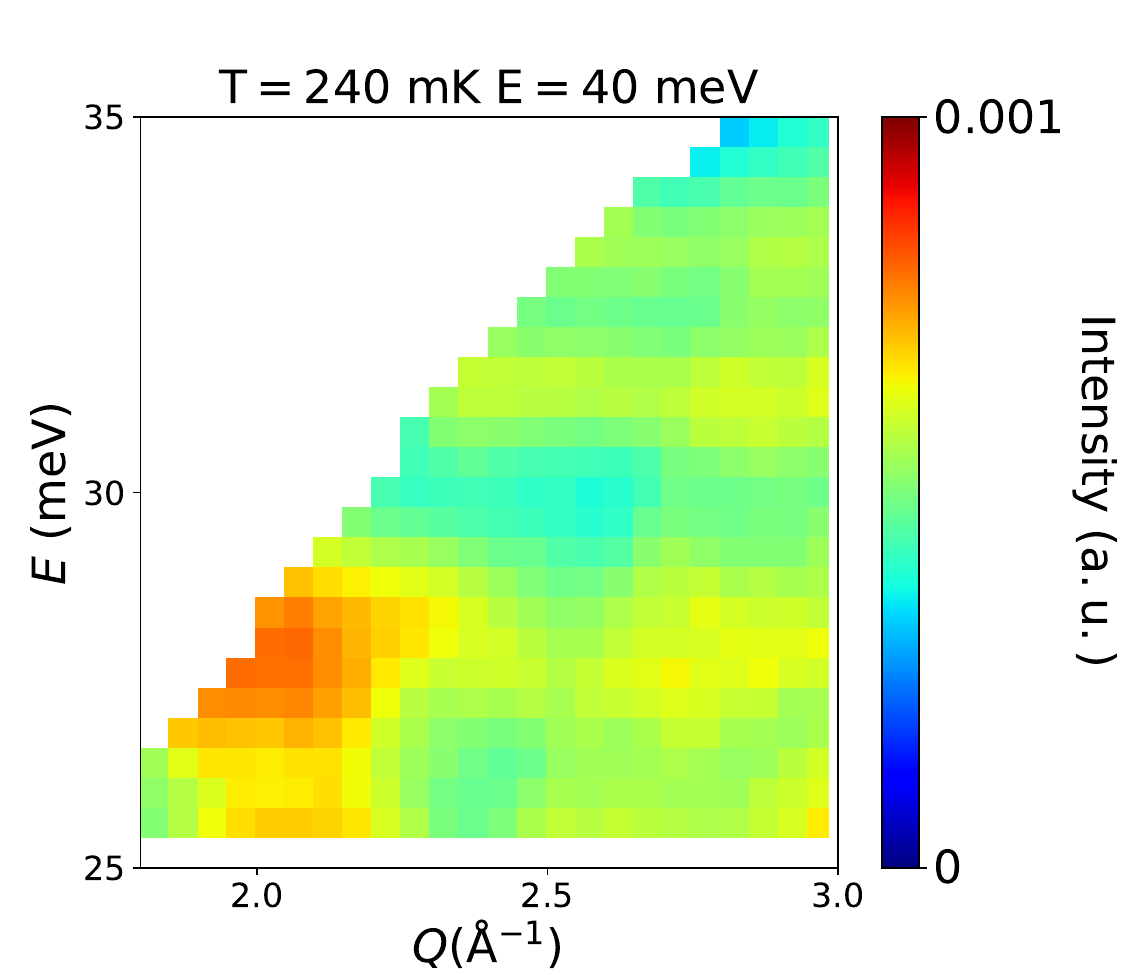}
\caption{Higher energy excitation spectrum for CoZnMo$_3$O$_8$.  The data are collected at T = 240 mK using the Fine-Resolution Fermi Chopper Spectrometer, SEQUOIA, SNS with incident neutron energy of E = 40 meV with the incident neutrons along the c axis. The E vs Q plot was plotted using powder average. Colour bars indicate scattering intensity in arbitrary units (a.u.).
\label{fig_CEF}
}
\end{figure}

As detailed in Ref.~\onlinecite{lines1963magnetic, winter2022magnetic}, the lowest crystal field excitations are expected to appear in the energy range of $E = \lambda_{\rm Co}/2 \approx 30$ meV, where $\lambda_{\rm Co} \approx 60$ meV is the atomic spin-orbit coupling strength of Co$^{2+}$. Due to a trigonal crystal field term $\Delta_2 < 0$ (according to notation in Ref.~\onlinecite{winter2022magnetic}), the excitations are expected to be split into two levels. In the higher energy excitation spectrum (Fig.~\ref{fig_CEF}), we observe several non-dispersive excitations in this energy range, which likely points to more than one distinct $O_h$ environment within the sample. This finding is compatible with a significant intersite disorder, as the presence or absence of randomly placed Co$^{2+}$ $T_d$ sites is likely to perturb slightly the local structures, leading to different environments for the $O_h$ Co$^{2+}$ ions. 

\section{Magnetic Couplings}

For the intralayer $O_h-O_h$ interactions, the computed interactions for the Z-bond are:
\begin{align}
\mathbb{J}_{||,Z} =\left(\begin{array}{ccc} 
0.341 & 0.141 & 0.112 \\ 
0.118 & 0.341 & 0.090 \\ 
0.090 & 0.112 & 0.315 
 \end{array}\right)
 \end{align}
 In the $J,K,\Gamma,\eta$ approximation, the best fit corresponds to $J_{||} = 0.53, K_{||} = -0.03, \Gamma_{||} = 0.05$ and $\eta = 0.66$. With these parameters:
 \begin{align}
\mathbb{J}_{||,Z}  \approx \left(\begin{array}{ccc} 
0.341 & 0.129 & 0.101 \\ 
0.129 & 0.341 & 0.101 \\ 
0.101 & 0.101 & 0.315
 \end{array}\right)
 \end{align}
With this choice, we ignore Dzyalloshinskii-Moriya interactions. 
For the interlayer $O_h - O_h$ interactions, the two sites are bisected by a common mirror plane. There are three types of interlayer bonds, which can be labelled X, Y, and Z in analogy with the intralayer couplings. For the interlayer ``Z-bond'', the mirror plane is formed by the cubic $z$-axis and $(x+y)$-axis. In this case, the computed coupling tensor is:
\begin{align}
\mathbb{J}_{\perp,Z} = \left(\begin{array}{ccc}
0.129 & -0.004 & 0.067 \\
-0.004 & 0.129 & 0.067 \\
0.063 & 0.063 & 0.199
\end{array} \right)
\end{align}
In the $J,K,\eta$ approximation, the best fit corresponds to $J_{\perp} = 0.32, K_\perp = 0.1, \Gamma_\perp = -0.17$, and $\eta = 0.52$. With these, 
\begin{align}
\mathbb{J}_{\perp,Z} \approx \left(\begin{array}{ccc}
0.129 & -0.004 & 0.065 \\
-0.004 & 0.129 & 0.065 \\
0.065 & 0.065 & 0.199
\end{array} \right)
\end{align}
For the $O_h - T_d$ couplings, we find some bond-dependence. Similar to the interlayer interactions, these bonds are bisected by a mirror plane, and can be labelled X, Y, and Z by the same convention. For the Z-bond, we compute:
\begin{align}
\mathbb{J}_{TO,Z} = \left(\begin{array}{ccc}
2.79 & -0.15 & 0.34 \\
-0.15 & 2.79 & 0.34 \\
0.47 & 0.47& 3.01
\end{array} \right)
\end{align}
As discussed in the main text, we approximate $K_{TO} = \Gamma_{TO} = 0$, and introduce pure Ising anisotropy. With this approximation, the best fit interaction tensor is (with $J_{TO} = 3.2, \eta = 0.83$):
\begin{align}
\mathbb{J}_{TO,Z} = \left(\begin{array}{ccc}
2.86 & 0.18 & 0.18 \\
0.18 & 2.86 & 0.18 \\
0.18 & 0.18& 2.86
\end{array} \right)
\end{align}
Finally, in order to further simplify the couplings, we took $\eta$ to be the same for every bond, and tuned $J_{TO}, \eta$, and $A_c$ to be consistent with the observed spinwave energies for Co$_2$Mo$_3$O$_8$. 

\section{heat capacity extrapolation}

To properly estimate the magnetic entropy $S_{mag}$ from specific heat, one needs to integrate the magnetic contribution to the specific heat $C_{mag}/T$ from absolute zero, which then requires appropriate extrapolation of the data from the lowest experimental temperature $\sim$ 50 mK. For zero field, $C_{mag}/T$ diverges as the temperature decreases. Such divergence could be either attributed to (1) hyperfine interactions due to large nuclear moment of Co and isotopes of Zn, or (2) impurity-induced Schottky anomalies. We first rule out (1) as the hyperfine interactions are enhanced as the magnetization increases (by applying the magnetic field), causing the divergence to occur at higher temperature \cite{Rai2018}. In CoZnMo$_3$O$_8$ the divergence of $C_{mag}/T$ is completely suppressed as the field of 2~T is applied, as shown in Fig.~\ref{fig_1}(g-h).

\begin{figure}[t]
\includegraphics[width=\columnwidth]{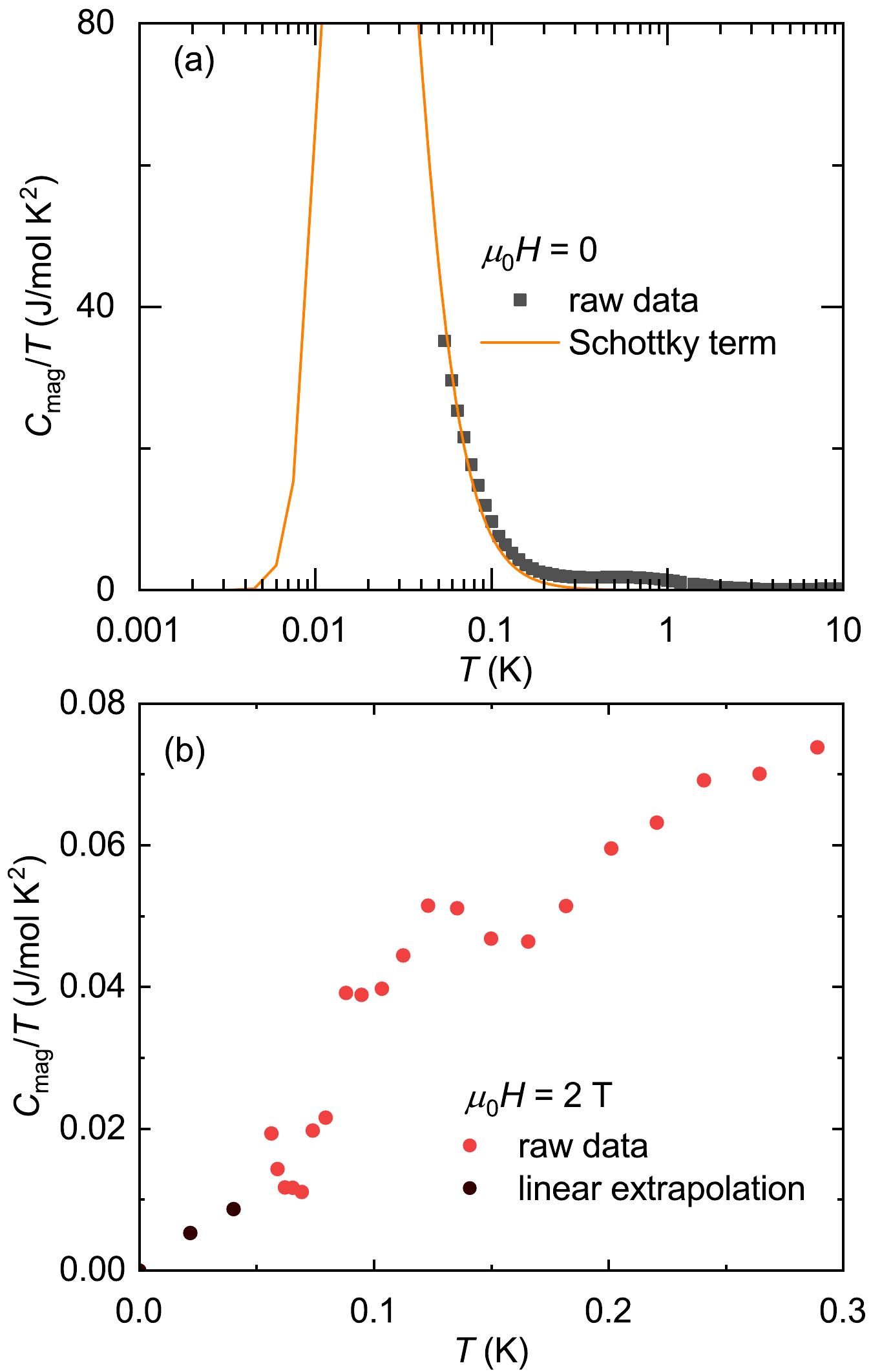}
\caption{Magnetic contribution to the specific heat $C_{mag}/T$ under zero field (a) and 2 T (b). An orange curve in (a) is a Schottky contribution to the specific heat from impurities. Black circles in (b) are extrapolated data from which $S_{mag,2T}$ at the field of 2 T is calculated and shown in Fig. 1(h). }
\label{fig_HC}
\end{figure}

Given the fact of considerable amount of impurities in CoZnMo$_3$O$_8$, (2) the induced Schottky anomaly could most probably be the origin of the divergence that the latter is a high-temperature tail of the former. We fitted the data using a two-level Schottky term with an energy splitting between two levels $\Delta$. It turns out with $\Delta = 0.06$~K, the Schottky term describes the data well well, as shown in Fig.~\ref{fig_HC}(a). When the field of 2~T is applied, the two levels are fully depopulated and hence the divergence of $C_{mag}/T$ is suppressed. By supplementing the Schottky term to the experimental data, we calculated the magnetic entropy $S_{mag}$, as shown in Fig.~\ref{fig_1}(h). $C_{mag}/T$ data measured at 2 T decrease almost linearly with temperature, and hence we extrapolated the data to absolute zero by linear extrapolation, as shown in Fig.~\ref{fig_HC}(b), and obtained $S_{mag,2T}$. Though the estimated $S_{mag,zero}$ and $S_{mag,2T}$ contain unavoidable error, the comparison between the two reflects an unambiguous fact that $S_{mag}$ reaches Rln2 at a much lower temperature in zero field than that in 2 T.

\end{document}